\documentclass[5p]{elsarticle}

\usepackage{amsmath,amssymb,epsfig,color}
\usepackage{graphicx}

\begin{document}

\title{Mode-by-mode fluid dynamics for relativistic heavy ion collisions}


\author{Stefan Floerchinger and Urs Achim Wiedemann}


\address{Physics Department, Theory Unit, CERN, CH-1211 Gen\`eve 23, Switzerland\\
                E-mail: stefan.floerchinger@cern.ch, urs.wiedemann@cern.ch}

\begin{abstract}
We propose to study the fluid dynamic propagation of fluctuations in
relativistic heavy ion collisions differentially with respect to their azimuthal, 
radial and longitudinal wavelength. To this end, we introduce a background-fluctuation
splitting and a Bessel-Fourier decomposition of the fluctuating modes. We 
demonstrate how the fluid dynamic evolution of realistic events can 
be build up from the propagation of individual modes. 
We describe the main elements of this mode-by-mode fluid dynamics,
and we discuss its use in the fluid dynamic analysis of heavy ion collisions.
As a first illustration, we quantify to what extent only fluctuations of sufficiently large 
radial wave length contribute to harmonic flow coefficients. We find that 
fluctuations of short wave length are suppressed not only due to larger dissipative 
effects, but also due to a geometrical averaging over the freeze-out hyper surface.  
In this way, our study
further substantiates the picture that harmonic flow coefficients give access to a
coarse-grained version of the initial conditions for heavy ion collisions, only. 
\end{abstract}


\maketitle

In nucleus-nucleus collisions at the LHC and at RHIC, the dependence of 
soft hadron spectra on transverse momentum, on azimuthal orientation, on 
centrality and on particle species is understood since recently as fluid dynamic 
response to fluctuating initial conditions~\cite{Alver:2010gr,Mishra:2007tw,Broniowski:2007ft,Sorensen:2008zk,Takahashi:2009na}, for reviews see 
ref.\ \cite{Heinz:2013th,Gale:2013da}. 
This success of a fluid dynamic description is significant mainly for two reasons. 
First, the high sensitivity of fluctuations to dissipative properties of the produced fluid implies that fluctuations are promising tools for constraining the transport properties 
of dense QCD matter with unprecedented accuracy\cite{Qiu:2011iv,Schenke:2011bn}. 
Second, since minimal dissipation implies maximal transparency to fluctuations, 
fluctuations in the initial stage of the collision can survive the time evolution. 
Therefore, the analysis of fluctuations may give access to the initial pre-equilibrium 
state and its fast evolution towards local 
equilibrium~\cite{Bhalerao:2011yg,Schenke:2012wb}. 

Motivated by these perspectives, many recent works have explored the dynamical 
relation between fluctuating initial conditions and experimentally accessible data. 
One important line of research is to characterize initial conditions for event averages and
event ensembles in terms of eccentricities or closely related cumulants of the
initial (entropy) density distribution~\cite{Teaney:2010vd,Teaney:2012ke}, 
and to propagate entire events in viscous fluid dynamic simulations to the hadronic 
final state~\cite{Schenke:2011bn,Holopainen:2010gz,Qiu:2011hf,Gardim:2011xv,Qian:2013nba,Petersen:2012qc,Deng:2011at,Bhalerao:2011bp,Gale:2012rq,Niemi:2012aj}.
To date, this approach provides the most detailed test for the validity
of a fluid dynamic description of heavy ion collisions. Despite this success of a 
cumulant-based characterization of initial conditions, several reasons motivate 
to explore alternative ones~\cite{Floerchinger:2013vua,Staig:2010pn,Staig:2011wj,Gubser:2010ui,Florchinger:2011qf,ColemanSmith:2012ka,Springer:2012iz}.  
First, it is a problem well-known in probability theory that
while any positive transverse density distribution can be characterized uniquely in terms
of its infinite set of moments or cumulants, it is not possible to find (beyond the cumulants that
determine a Gaussian) a positive density configuration
corresponding to a finite set of cumulants such that all higher ones vanish. 
Strictly speaking, this implies that single cumulants cannot be propagated in fluid dynamics. 
Second, it is unknown how to extend a cumulant
expansion to vector and tensor fields, as is needed e.g. if one wants to explore the
natural possibility that fluctuations are manifest not only in the initial densities but
also in the velocity field and shear viscous tensor. Finally, each cumulant receives typically 
contributions from fluctuations on various different wavelengths. There are advantages
in decomposing initial fluctuations in an orthonormal basis of modes, but such
bases have been used so far only in studies that formulate fluid dynamic perturbations
on top of simple analytically known background fields with extended symmetries~\cite{Staig:2010pn,Staig:2011wj,Gubser:2010ui,Florchinger:2011qf,Springer:2012iz}. 

In a compagnon work~\cite{Floerchinger:2013vua}, we have discussed how to 
characterize initial conditions in an orthornormal basis for scalar
densities as well as vector and tensor fluctuations, constructed such that single 
fluctuating modes define positive densities and can therefore be propagated 
individually, mode-by-mode. In the present letter, we provide the first application 
of such a mode-by-mode fluid dynamics to realistic initial conditions, equation of state
and transport properties. More specific, we characterize event samples as 
probability distributions over a basis of modes, we propagate each mode 
individually, and we build up experimental observables as superpositions of the 
individually propagated modes. We comment on how this can improve our understanding
of why specific fluctuations in the initial state survive or do not survive the dynamic
evolution.

\begin{figure}
\includegraphics[width=0.24\textwidth]{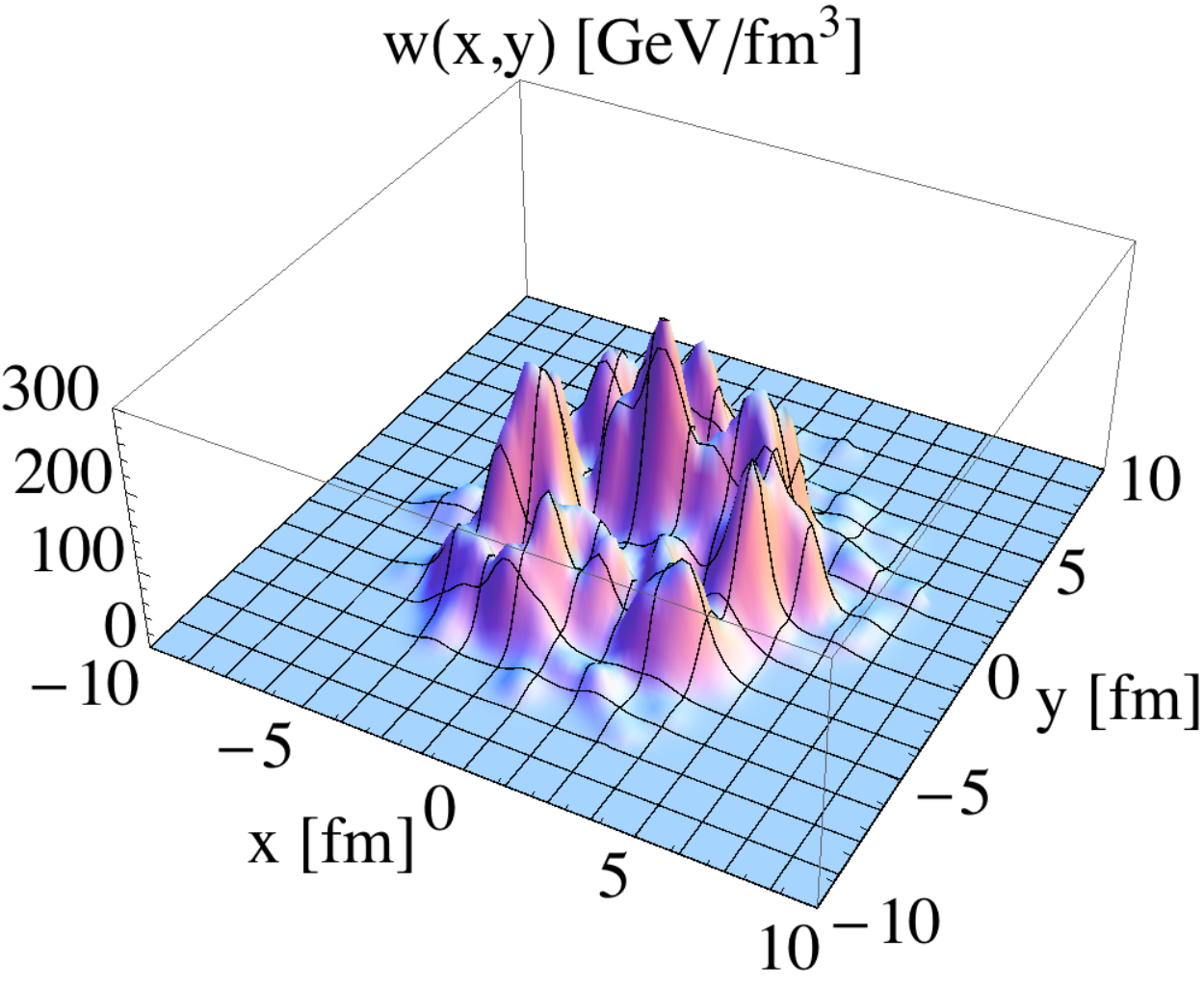}\hfill
\includegraphics[width=0.24\textwidth]{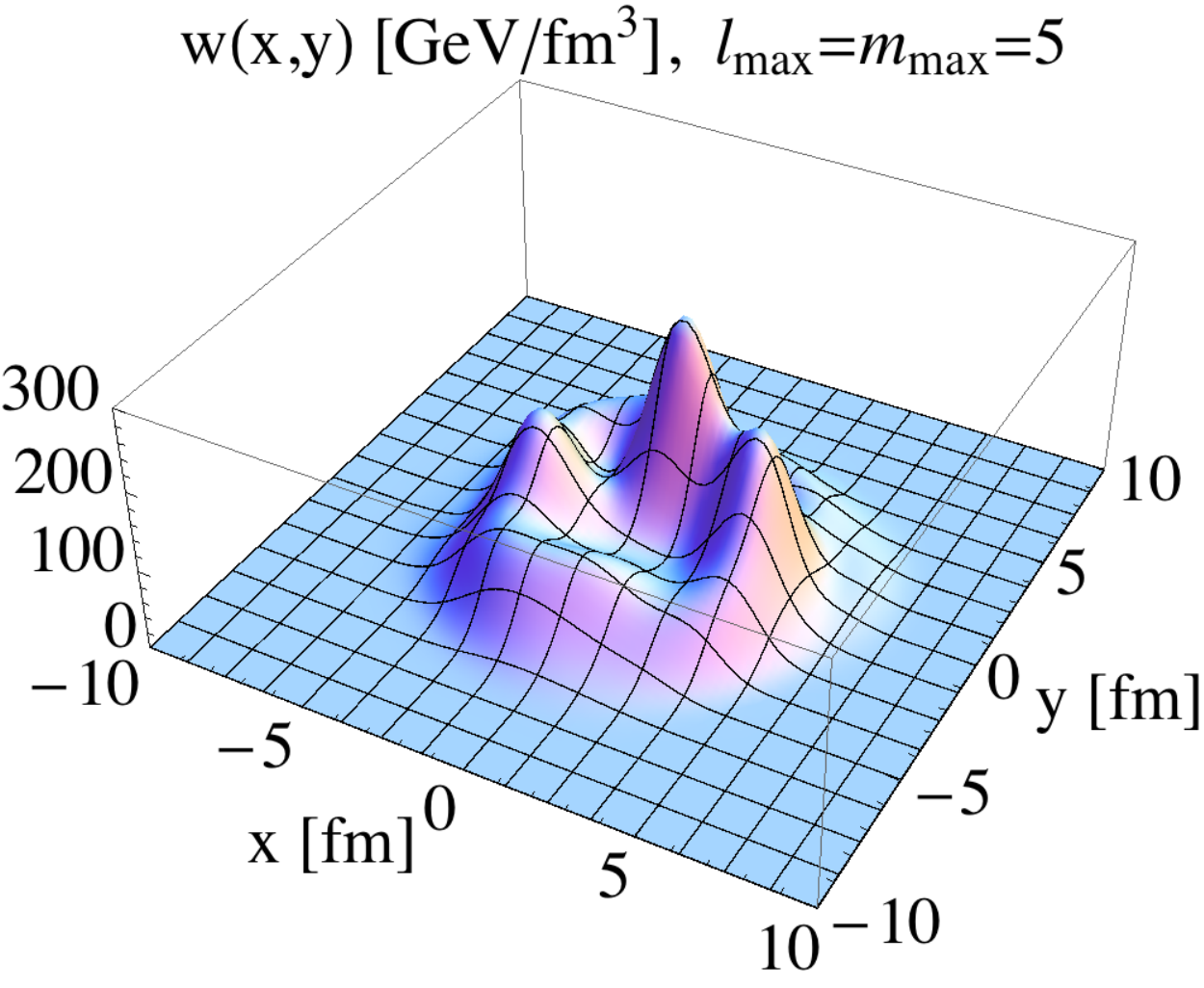}
\includegraphics[width=0.24\textwidth]{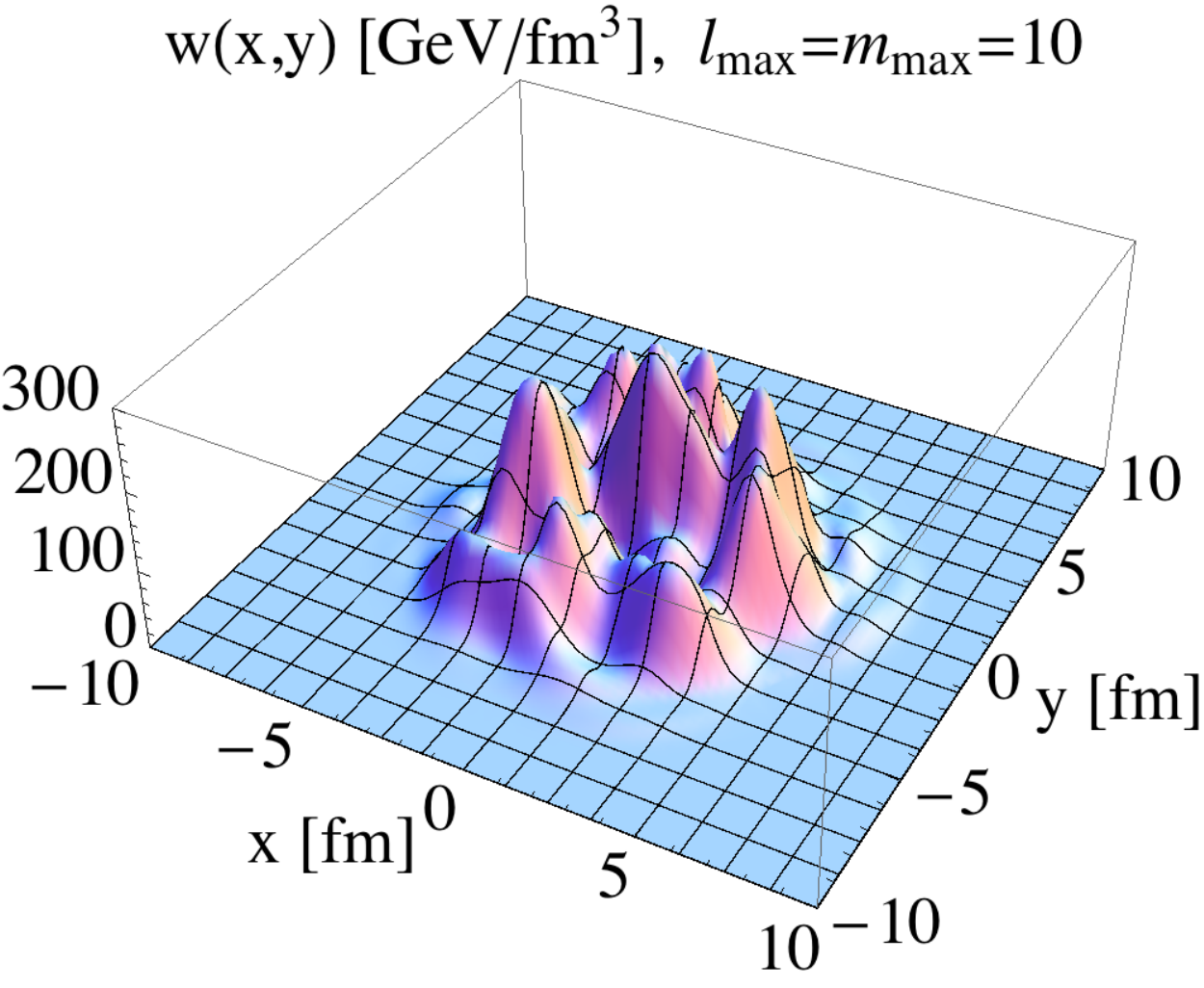}\hfill
\includegraphics[width=0.24\textwidth]{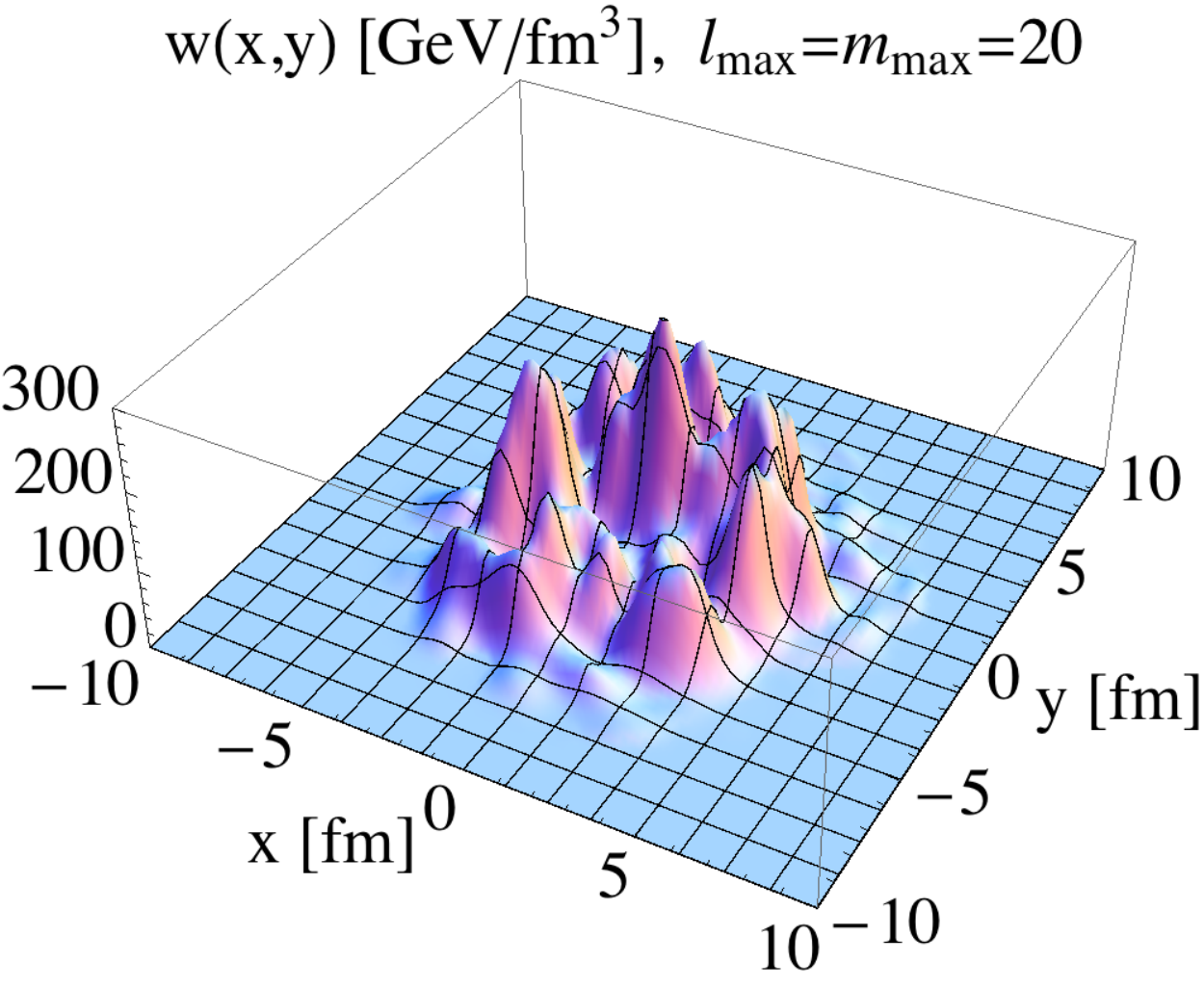}
\caption{Initial transverse enthalpy density $w$ of the MC Glauber model of Ref.~\cite{Holopainen:2010gz}.
Contributions of single participants are smeared by Gaussians with $\sigma = 0.4$ fm and reweighted by the
number of collisions according to Ref.~\cite{Qiu:2011hf} ($x_{\rm coll} = 0.118$). A finite number of
modes $(m_{\rm max},l_{\rm max})$ in \eqref{eq:BesselFourierExpansion1}, \eqref{eq:BesselFourierExpansion2} 
allows one to reconstruct $w$ efficiently.
}
\label{fig1}
\end{figure}

\paragraph{Initial conditions}
The dynamical evolution is initialized by specifying fluid dynamic fields on some hyper-surface, usally taken at fixed
$\tau=\sqrt{t^2-z^2}=\tau_0$. 
Most generally, a model of the initial conditions is then defined in terms of a (functional) probability distribution $p_{\tau_0}$
of the energy density $\epsilon$ or enthalpy $w=\epsilon+p$, the fluid velocity $u^\mu$, the shear stress $\pi^{\mu\nu}$, 
the bulk viscous pressure $\pi_\text{bulk}$ and possibly other fluid dynamic fields at $\tau_0$
\begin{equation}
p_{\tau_0}[w, u^\mu, \pi^{\mu\nu}, \pi_\text{bulk}]\, .
\label{eq:FunctionalProbabilityDistribution}
\end{equation}
To discuss properties of the distribution $p_{\tau_0}$, we focus on one fluid dynamic
field only, the enthalpy $w$. We consider fluctuations around a smooth background field $w_{\rm BG}$ that is boost invariant and
azimuthally symmetric. Longitudinal and azimuthal fluctuations in $w$ are characterized by a standard Fourier expansion,
\begin{eqnarray}
&& w(\tau_0,r,\phi,\eta) - w_{\rm BG}(\tau_0,r) \nonumber \\
&& \quad = 
\int \frac{d k_\eta}{2\pi} \sum_{m=-\infty}^\infty \, e^{ik_\eta \eta+ i m\phi}\, {w}^{(m)}(\tau_0,r,k_\eta) \, .
\label{eq:BesselFourierExpansion1}
\end{eqnarray}
For notational simplicity, we assume in what follows longitudinal boost invariance, i.e., we neglect any dependence on 
$k_{\eta}$. If needed, our discussion is extended easily to the case of a non-trivial $k_\eta$-dependence. 
The radial dependence is expanded in terms of Bessel functions $J_m$ that have appropriate
boundary conditions at $r=0$, 
\begin{equation}
w^{(m)}(\tau_0,r) = w_{\rm BG}(\tau_0,r)\, \sum_{l=1}^\infty J_m\left(k^{(m)}_l r\right) \tilde{w}^{(m)}_l.
\label{eq:BesselFourierExpansion2}
\end{equation}
Here, the radial wave vectors $k_l^{(m)} = z^{(m)}_l/R$ are set by the $l$-th zeroes $z^{(m)}_l$ of $J_m(z)$ and an overall scale $R$ ($R=8$ fm for the results presented here). The main difference between (\ref{eq:BesselFourierExpansion2}) and the
Bessel-Fourier expansion proposed first in~\cite{ColemanSmith:2012ka} is that 
we include the normalization factor $w_{\rm BG}(\tau_0,r)$ on the right hand side. 
This ensures that the enthalpy density is positive everywhere even when only one 
or a few of the coefficients $\tilde{w}^{(m)}_l$ are non-vanishing. 
The azimuthal and radial wavenumber $m$ and $l$ can be restricted to the ranges 
$(-m_{\rm max},\ldots,m_{\rm max})$ and $(1,\ldots,l_{\rm max})$ when the spatial resolution is bound. Lemoine's discrete
Bessel transformation provides a CPU-inexpensive method for determining $\tilde{w}^{(m)}_l$~\cite{Lemoine,Floerchinger:2013vua}.
Fig.\ \ref{fig1} illustrates 
for a phenomenologically relevant enthalpy density that fluctuations in a single event can be characterized satisfactorily in 
terms of a small set of $m_{\rm max}=l_{\rm max}\simeq O(10)$ Bessel coefficients $w^{(m)}_l$ in \eqref{eq:BesselFourierExpansion1}, \eqref{eq:BesselFourierExpansion2}. 

Event samples can have {\it statistical} symmetries that are broken by event-wise fluctuations. For instance, a sample at vanishing impact parameter $b=0$ 
will have statistical azimuthal symmetry. In this case, we choose to identify the background field in \eqref{eq:BesselFourierExpansion1} with the event average $w_{BG} \equiv \langle w \rangle$. Also at finite $b$, it can be advantageous to choose an azimuthally symmetric  $w_{BG}$ even though this symmetry is not realized statistically; one can define e.g. $w_{BG}$ as the average over
azimuthally randomized events. The azimuthal dependence of the event sample is then encoded in the 
event-averaged Bessel coefficients $\bar{w}^{(m)}_l = \langle \tilde{w}^{(m)}_l \rangle$ that can take
non-zero values for even integers $m$ when $b \not=0$. 
We have tested in model studies that the ansatz \eqref{eq:BesselFourierExpansion1}, \eqref{eq:BesselFourierExpansion2} 
is as accurate for classes of semi-peripheral collisions, as for central ones~\cite{Floerchinger:2013vua}.

\begin{figure}
\includegraphics[width=0.4\textwidth]{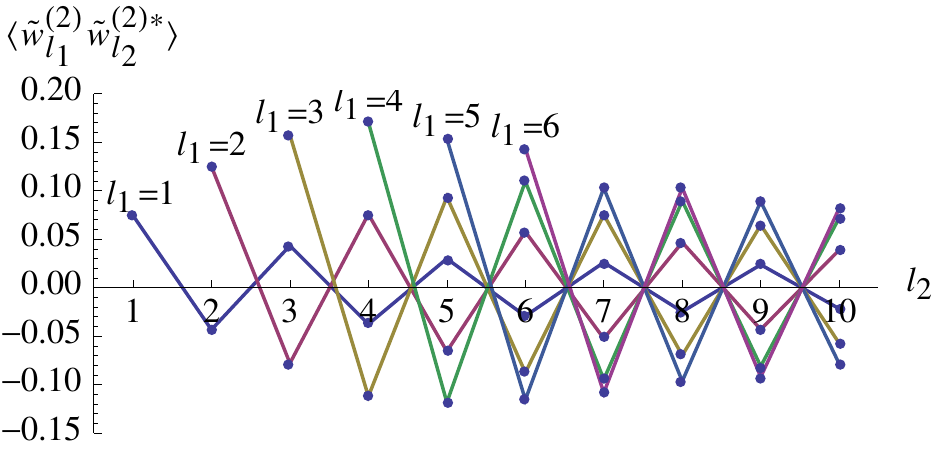}
\caption{Correlation of Fourier-Bessel components of the enthalpy density $\langle \tilde w^{(m)}_{l_1} \tilde w^{(m)*}_{l_2} \rangle$ according to the Monte-Carlo Glauber model for central collisions. We plot this for $m=2$, different values of $l_1$ and as a function of $l_2$. The curves look similar for $m=1$ or $m=3$.}
\label{fig2}
\end{figure}

Since the coefficients $\tilde{w}^{(m)}_l$ characterize 
single events fully, the functional probability density $p_{\tau_0}$ becomes a function of
a set of numbers $\tilde{w}^{(m)}_l$. We have established~\cite{Floerchinger:2013vua} 
that for currently used models of initial conditions, $p_{\tau_0}$ satisfies to good approximation the properties
of a Gaussian probability distribution. For $b=0$,
\begin{equation}
p_{\tau_0} = \frac{1}{{\cal N}} 
\exp {\bigg [} - \frac{1}{2}\sum_{m=-m_\text{max}}^{m_\text{max}} \sum_{l1,l2=1}^{l_\text{max}} T^{(m)}_{l_1 l_2} \tilde w^{(m)*}_{l_1}  \tilde w^{(m)}_{l_2} {\bigg ]}\, .
\label{eq:probabilityDistributionGaussian}
\end{equation}
Thus, $p_{\tau_0}$ is fully characterized in terms of $w_\text{BG} \equiv \langle w \rangle$ 
and the two-point correlators
\begin{equation}
(T^{(m)})^{-1}_{l_1 l_2} = \langle  \tilde w^{(m)}_{l_1}  \tilde w^{(m)*}_{l_2} \rangle\, .
\label{twopoint}
\end{equation}
Fig.~\ref{fig2} shows the $m=2$
two-point correlators (\ref{twopoint}) for the Monte Carlo Glauber model described 
in~\cite{Holopainen:2010gz,Floerchinger:2013vua}. 
From these data (for all $m< m_{\rm max}$), event samples of initial conditions can be generated easily.
Since a mode $\tilde{w}^{(m)}_l$ corresponds 
to a radial wavelength $1/k_l^{(m)} = R/z^{(m)}_l$ that decreases with increasing $l$,  
Fig.~\ref{fig2} shows how fluctuations on different radial length scale decorrelate as they are separated in scale. 
%
%

\paragraph{Dynamic evolution}
The above classification of initial conditions introduces naturally a background-fluctuation splitting 
$w=w_{\rm BG}+w_{\rm F}$, $u^\mu=u^\mu_{\rm BG}+u^\mu_{\rm F}$ etc. of all fluid fields. Instead of solving the
relativistic fluid dynamic equations for the fields $w$, $u_\mu$ etc. event-by-event, we solve for the smooth
non-fluctuating background field once and for all, and we propagate the full basis of initial fluctuating modes
with wave-numbers $(l,m)$ as perturbations on this background field.

\begin{figure}
\centering
\includegraphics[width=0.4\linewidth]{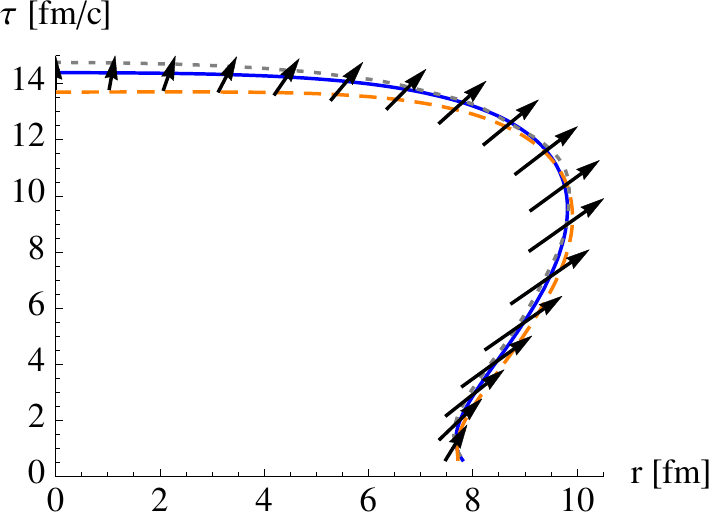}\hfill
\includegraphics[width=0.24\textwidth]{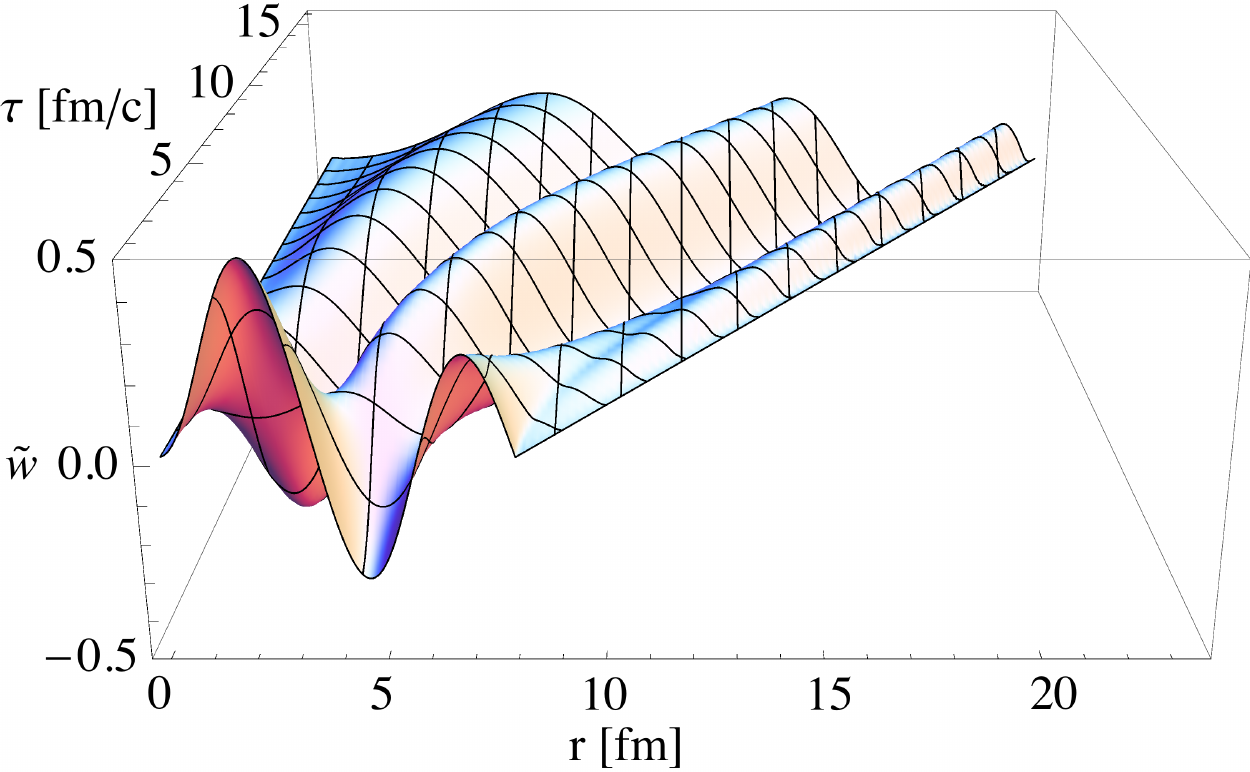}
\includegraphics[width=0.24\textwidth]{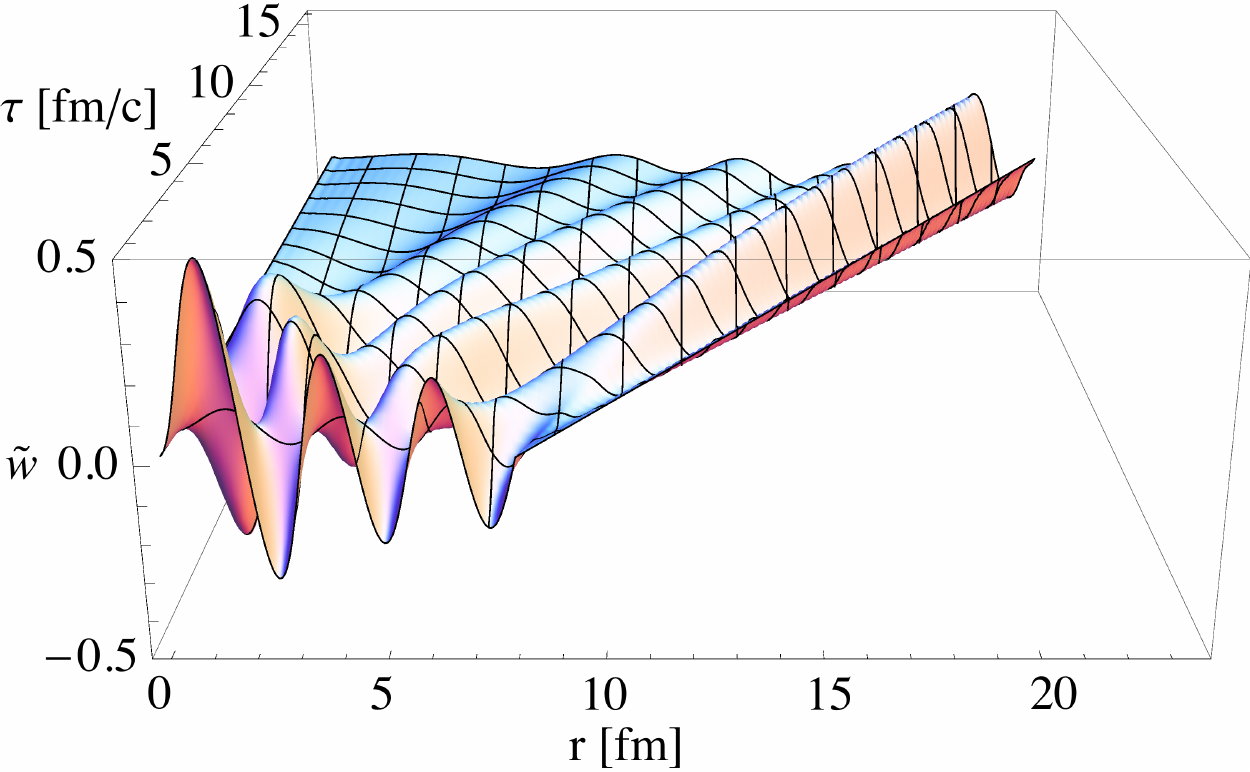}\hfill
\includegraphics[width=0.24\textwidth]{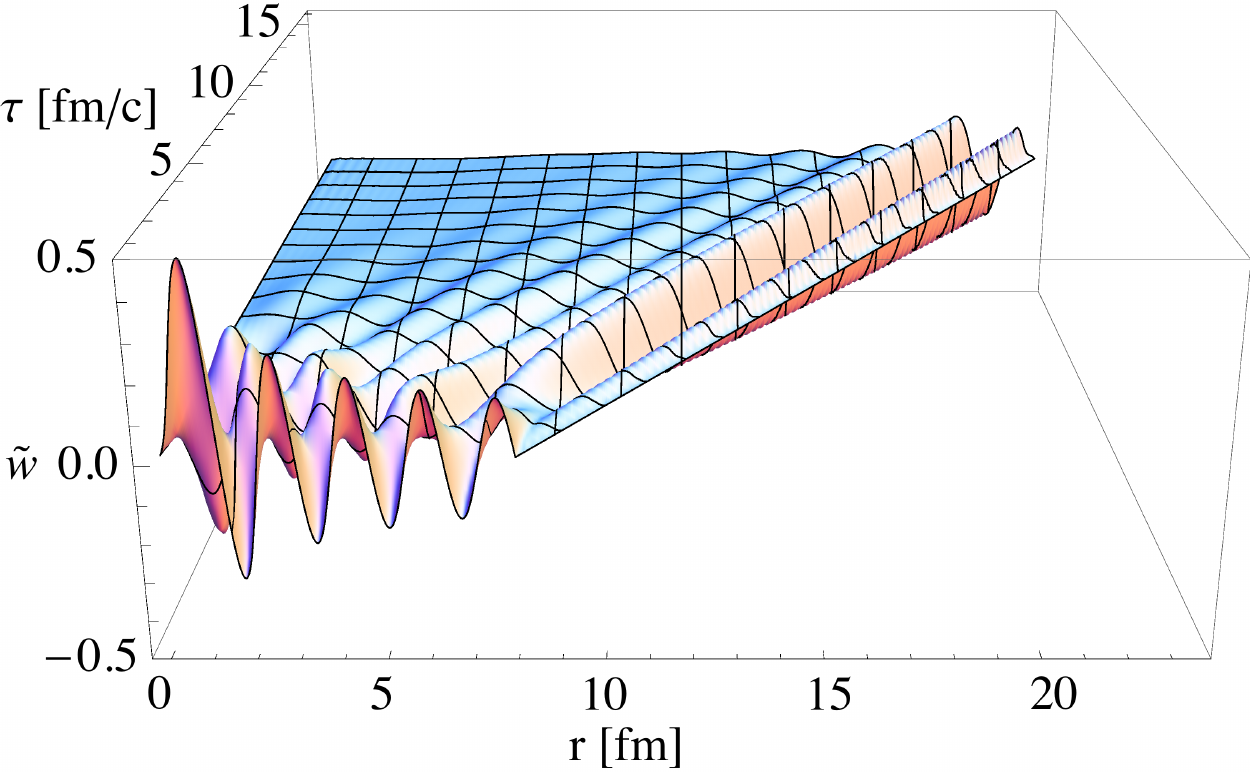}
\caption{Upper left: Freeze out curve of event-averaged background field  for a central Pb-Pb collision at LHC energy 
at $T_\text{fo}=130 \,\text{MeV}$ and for different choices of the shear viscosity to entropy ratio: $\eta/s=0$ (dotted), 
$\eta/s=0.08$ (solid) and $\eta/s=0.3$ (dashed). The arrows indicate the direction of the fluid velocity at freeze-out for 
the case $\eta/s=0.08$. Upper right and lower panel: Time evolution of fluctuations in the normalized enthalpy density 
$\tilde{w} = \delta w / w_{\rm BG}$ for $\eta/s = 0.08$ and three different modes of initial conditions corresponding to 
azimuthal wave number 
$m=2$ and radial wave numbers $l=3$,  $l=6$ and $l=9$, respectively. 
At large $r$ where 
$w_{\rm BG}(r,\tau)$ is small, small fluctuations $\delta w$ can be visually prominent in $\tilde{w} = \delta w / w_{\rm BG}$.}
\label{fig3}
\end{figure}

Relativistic viscous fluid dynamic solutions of event-averaged background fields are well-documented. 
We follow ref.\ \cite{Qiu:2011hf} in using the equation of state s95p-PCE which combines lattice QCD results at high temperatures with a hadron resonance gas at low temperatures. It implements also a chemical freeze-out at $T=165 \,\text{MeV}/k_B$. The default value of the shear viscosity to entropy ratio is  $\eta/s=0.08\, \hbar/k_B$ and the corresponding relaxation time $\tau_\text{Shear}=0.23\,\hbar/(k_B T)$. The evolution is initialized at $\tau_0=0.6\,\text{fm/c}$ with initial flow and shear stress fields corresponding to the Navier-Stokes form of a longitudinal Bjorken expansion. The background enthalpy is initialized 
as $w_{\rm BG} = \langle w \rangle$ at $b = 0$. The entropy in both background and fluctuations scale with $(1-x)N_{\rm part}/2 + x\, N_{\rm coll}$, $x = 0.118$, as in 
ref.~\cite{Qiu:2011hf}. Fig.~\ref{fig3} shows the freeze-out curves resulting from
fluid dynamic evolution of this azimuthally symmetric background field. 
They are consistent with published benchmarks.

The evolution equations for the fluctuations $w_F$, $u_F^\mu$ etc.\ depend on the solution for the background fields.
They become particularly simple if treated as small perturbations that can be linearized. 
For a given Fourier mode specified by $m$ and $k_\eta$, the evolution equation becomes then $1+1$ 
dimensional and with the Bessel expansion as in eq.\ \eqref{eq:BesselFourierExpansion2} it reduces for all $\tau$ to 
a set of coupled ordinary differential equations which we solve numerically. This set-up extends the strategy of
Refs.~\cite{Staig:2010pn,Florchinger:2011qf} to arbitrary background fields and arbitrary classes of {\it initial} fluctuations,
including initial fluctuations in the fluid velocity~\cite{Florchinger:2011qf} or shear. In fig.~\ref{fig3},  we show for the 
normalized enthalpy density $\tilde{w} = \delta w / w_{\rm BG}$ the spatial evolution for three modes of different radial 
wave number $l$. One sees that the viscous damping increases significantly for shorter radial wave-length, thus illustrating the importance of studying the effect of fluctuations differentially in $l$. We also find that the
viscous damping seen in fig.~\ref{fig3} increases strongly with $\eta/s$. On the other hand, modes with larger $l$ 
lead to more strongly oscillating patterns on the freeze-out surface and have therefore less impact on particle spectra even for $\eta/s = 0$. 

\paragraph{Freeze-out and particle spectra.} 
Hydrodynamics ceases to apply when interaction rates become too small to maintain  
local kinetic equilibrium. We assume that this happens when the background field drops below $T_\text{fo}=120\,\text{MeV}$. Particle distributions then freeze out. We determine them using the standard Cooper-Frye prescription neglecting resonance decays and hadronic rescatterings after freeze-out. (In principle these effects could be incorporated by solving the corresponding kinetic equations for the background and, in linearized form, for the fluctuations.)
The occupation numbers on the freeze-out surface are taken to be of ideal gas form with chemical potentials according to the equation of state 
s95p-PCE. Viscous corrections due to shear stress are approximated by the quadratic 
ansatz \cite{Teaney:2003kp}. 
Due to its azimuthal rotation invariance, the background field contributes to the $\phi$- and $y$-independent part of the one-particle spectrum $S(p_T)=dN/(p_T dp_T d\phi dy)$, only. If fluctuations on top of this background are small enough, 
their effect on particle spectra can be linearized,
\begin{equation}
\ln \left( \frac{dN^\text{single event}}{p_T dp_T d\phi dy} \right)\\
= \ln S_0(p_T) + \sum_{m,l} \tilde{w}^{(m)}_l e^{i m \phi} \theta^{(m)}_l(p_T)\, .
\label{onespectrum}
\end{equation}
Here, the functions $\theta^{(m)}_l(p_T)$ determine how the fluctuating modes of wave-numbers $(m,l)$ contribute to the
hadronic spectrum. In general, the $\theta^{(m)}_l$ depend also on rapidity and particle species. They are calculated as follows: The linearized hydrodynamical evolution equations on top of the background solution are solved for the initial condition corresponding to the mode $(m,l)$ in enthalpy density. All fluid fields resulting from this initialization are determined on the freeze-out surface and the corresponding contribution to the particle spectrum is determined from an appropriate linearization of the Cooper-Frye formula. Dividing finally by the background contribution to the particle spectrum yields $\theta^{(m)}_l(p_T)$.

One-particle spectra of event samples are obtained by averaging \eqref{onespectrum} over the probability distribution 
$p_{\tau_0}(\{ \tilde{w}^{(m)}_l \} )$.  
In close analogy, the calculation of two-particle correlations requires knowledge about the  
initial correlations between pairs of modes $\tilde w^{(m)}_{l_1} \tilde w^{(m)*}_{l_2}$ whose contribution
to the hadronic two-particle spectra is then determined by the product $\theta^{(m)}_{l_1}(p_{T}^a) \theta^{(m)}_{l_2}(p_{T}^b)$.
The double differential harmonic flow coefficient for event samples reads then to lowest order in $\tilde w^{(m)}_{l}$
\begin{equation}
v_m^2\{2\} (p_{T}^a, p_{T}^b) = \sum_{l_1,l_2=1}^{l_\text{max}} \theta^{(m)}_{l_1}(p_{T}^a) \theta^{(m)}_{l_2}(p_{T}^b) \langle \tilde w^{(m)}_{l_1} \tilde w^{(m)*}_{l_2} \rangle\, .
\label{harmflow}
\end{equation}
The single differential harmonic flow coefficients $v_m(p_{T})$ can be obtained from (\ref{harmflow})
 as appropriately weighted $p_T$ integrals. Note that in close analogy to the experimental procedure of extracting harmonic flow coefficients, \eqref{harmflow} does not invoke knowledge of a reaction plane but determines $v_m$ from the 
azimuthal dependence of two-particle correlations that have their dynamic origin in the azimuthal
correlations $\langle \tilde w^{(m)}_{l_1} \tilde w^{(m)*}_{l_2} \rangle$ between different fluctuating modes. 
In this way, once the functions $\theta^{(m)}_{l}$ are calculated for a given smooth background and the finite set 
of wave numbers $(m,l)$, the fluid dynamic propagation of arbitrary samples of fluctuations 
$p_{\tau_0}(\{ \tilde{w}^{(m)}_l \} )$ can be studied by simple matrix multiplication, see \eqref{onespectrum}, \eqref{harmflow}.
We note that this formulation assumes a linear relation between fluctuating modes at $\tau_0$ and hadronic distributions
at freeze-out. One could test the accuracy of such a linear relation
by comparing for selected events to results from full fluid dynamic simulations. This may also allow to identify characteristic
signatures of non-linear fluid dynamic behavior in heavy ion collisions which would be interesting in itself. 

Another possibility to estimate the effects of non-linear terms in the hydrodynamical evolution and at freeze-out is to treat them order-by-order in a perturbative expansion. The leading order in this perturbation theory for small fluctuations around a smooth but dynamically evolving background is the linear order presented in this paper. At next-to-leading or quadratic order one can study for instance how an $m=2$ and an $m=3$ mode interact and how this contributes to a signal for $v_5$. A more detailed discussion of this kind of perturbative treatment is left for a future publication.

\begin{figure}
\includegraphics[width=0.35\textwidth]{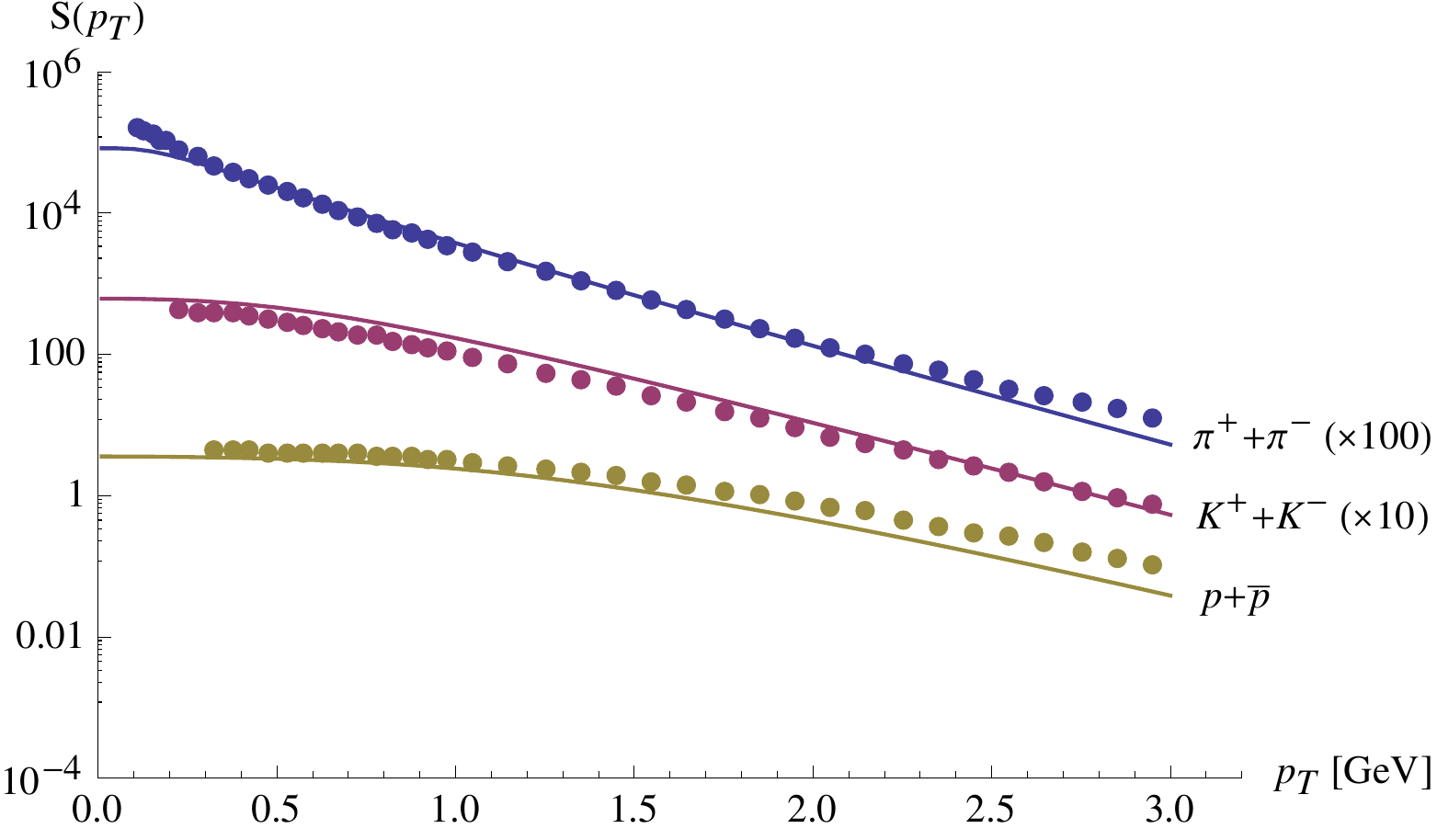}
\includegraphics[width=0.35\textwidth]{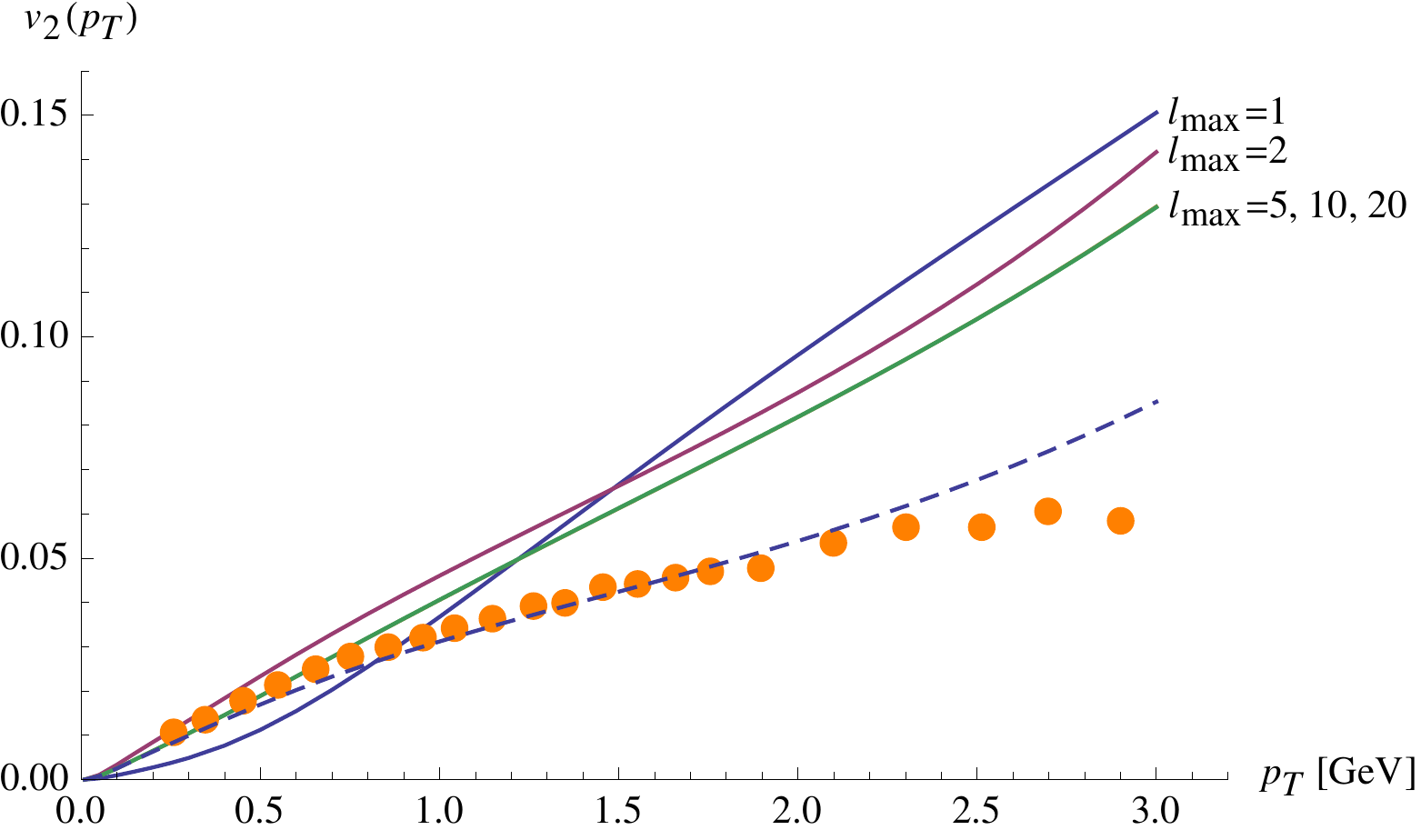}
\includegraphics[width=0.35\textwidth]{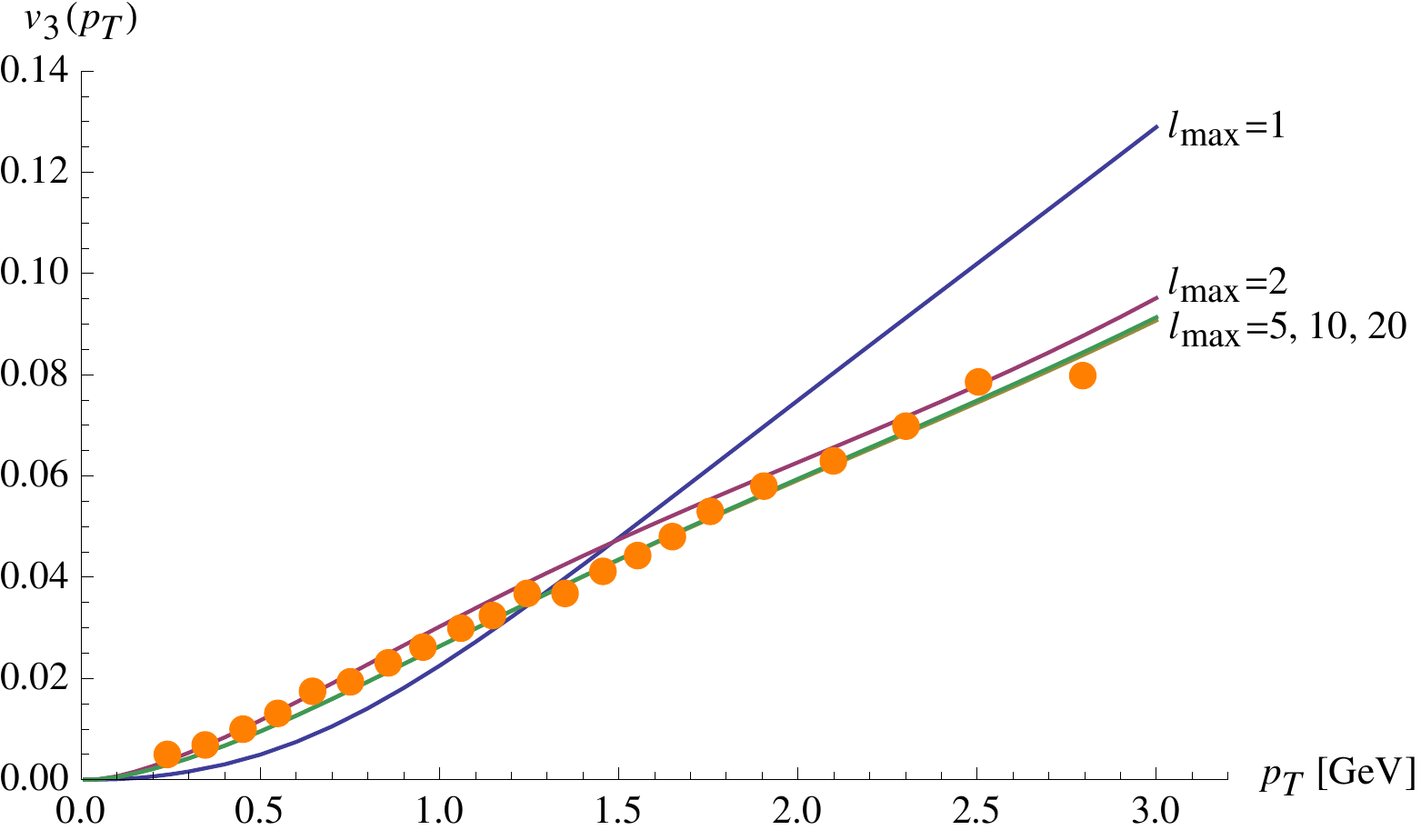}
\includegraphics[width=0.35\textwidth]{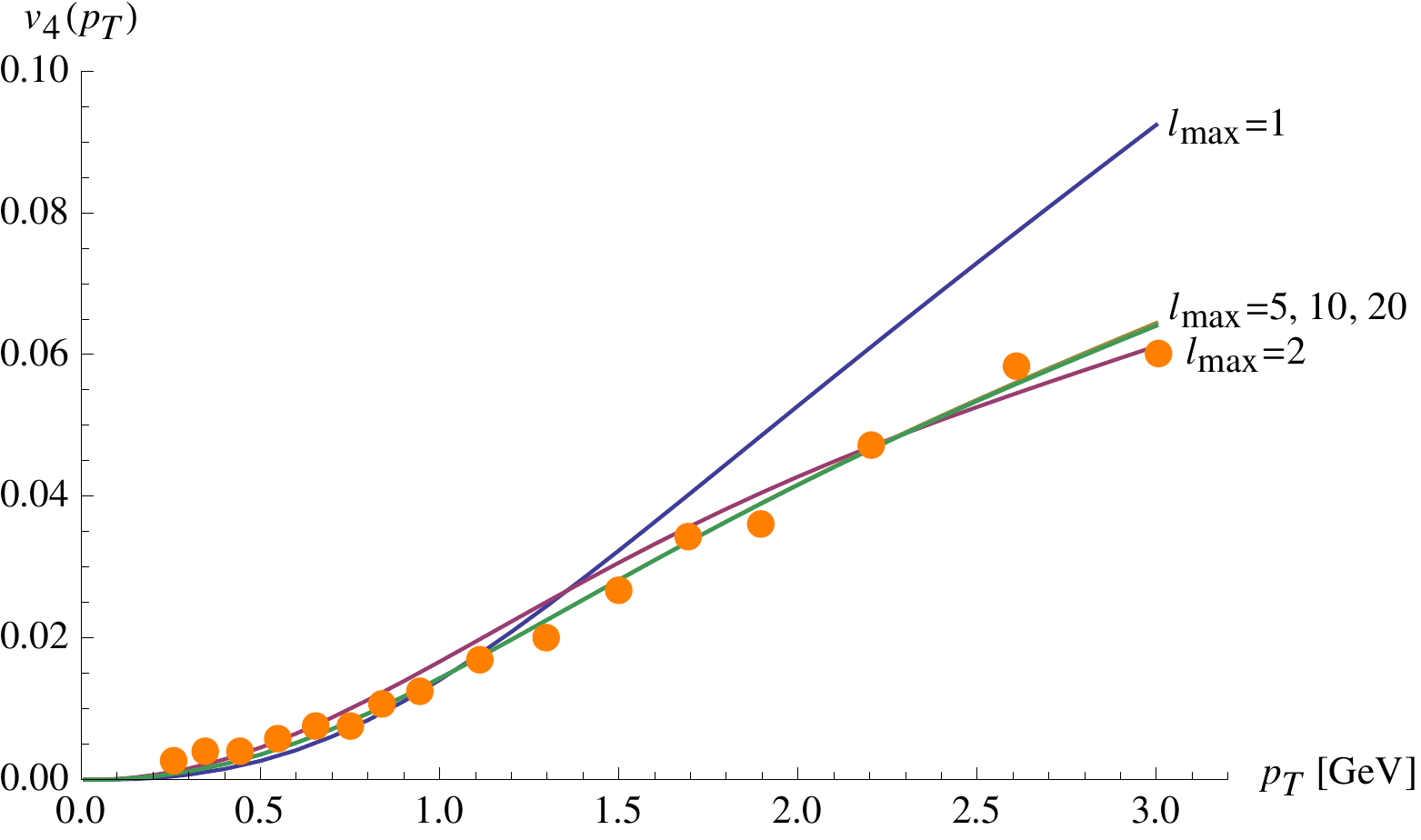}
\includegraphics[width=0.35\textwidth]{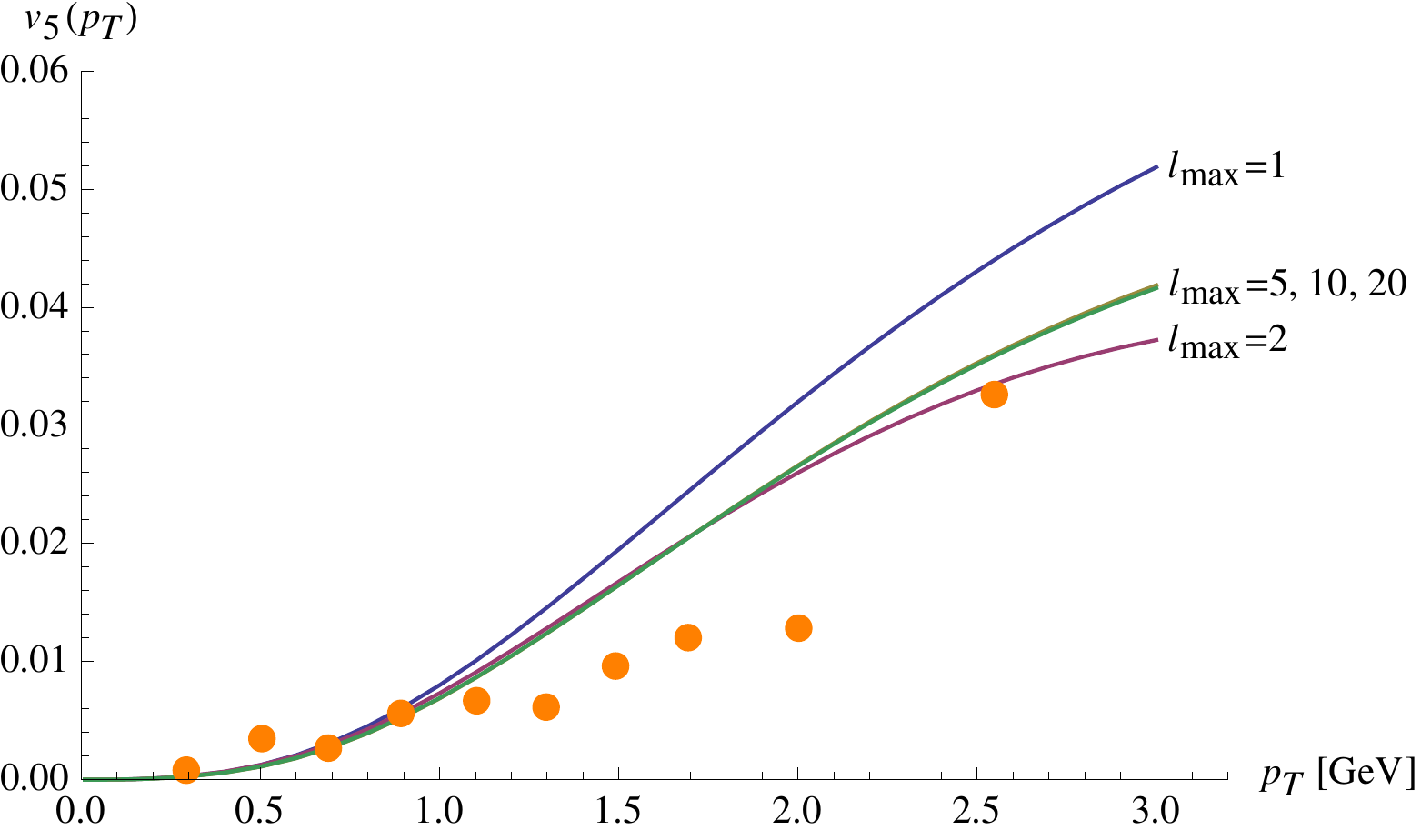}
\caption{Comparison of data on central Pb+Pb collisions at $\sqrt{s_{\rm NN}} = 2.76$ TeV
to fluid dynamic simulation described in the text. Upper panel: 
Particle spectra for pions, kaons and protons $S(p_T)=dN/(2\pi p_T d p_T dy)$, calculated from
(\ref{onespectrum}) and compared to data on $5\%$ most central events~\cite{Abelev:2012wca}.
Following panels: Elliptic ($v_2$), triangular ($v_3$), 4-th and 5-th order flow of charged particles, calculated for contributions from  $\pi^\pm, K^\pm, p$ and $\bar p$  from (\ref{harmflow}) with
$l_{\rm max} = 1,2,5,10,20$ and  compared to data on $2\%$ most central events. 
The dashed curve in the plot of $v_2$ shows results (for $l_{\rm max} = 20$) for a modified 
event sample in which  the weight of one particular fluctuating mode, $\tilde{w}_1^{(2)}$ is 
decreased by a factor $0.7$.}
\label{fig4}
\end{figure}

In fig.\ \ref{fig4}, we compare calculations of hadronic spectra and flow coefficients from mode-by-mode hydrodynamics to data for central Pb+Pb collisions taken by the ALICE Collaboration~\cite{Abelev:2012wca,ALICE:2011ab}. For the one-particle spectra 
of pions, kaons and protons, we find that fluctuations make a very small contribution to (\ref{onespectrum}), so that the model results shown in Fig.~\ref{fig4} are very close to
the result for an initial condition defined by the smooth background field without 
fluctuations. For the evolution of a smooth background field, we have checked 
on the level of the freeze-out hyper surface and on the level of spectra
that our simulations are quantitatively consistent with the hydrodynamic benchmarks established by the TECHQM Collaboration. Our study does not take into account
resonance decays and hadronic rescatterings after freeze-out (we do not switch to a hadron cascade model), and we made no attempt to improve agreement with data by optimizing input parameters of the fluid dynamic simulation
such as $\eta/s$ or the equilibration time $\tau_0$. It is therefore no surprise that one sees
some differences between data and numerical results for one-particle spectra. 
However, these one-particle spectra are sufficiently well reproduced to serve
as baseline for the present study, whose purpose is to illustrate how in 
a mode-by-mode fluid dynamic analysis, results on elliptic, triangular, 4-th and 5-th order 
flow are built up in terms of the contributions of individual fluctuations of characteristically
different radial wavelength. 
With this respect, our main conclusion is that the sum (\ref{harmflow}) converges quickly,
for small radial wave numbers $l \leq l_{\rm max} \approx 5$. This means that only fluctuations
of sufficiently large radial wavelength matter for the dynamics of flow coefficients. For the density distribution of the specific event shown in fig.\ \ref{fig1}, for instance, it is then only the 
coarse-grained information shown for $l_{\rm max} = m_{\rm max} = 5$ in the upper right 
panel of fig.\ \ref{fig1} that affects the value of flow harmonics in fig.\ \ref{fig4}. Since we observe
this rapid convergence in $l$ for minimal dissipative effects ($\eta/s = 1/4\pi$), we expect
this finding to be more general than the specific model study in which we have established it
here. 

\begin{figure}
\includegraphics[width=0.35\textwidth]{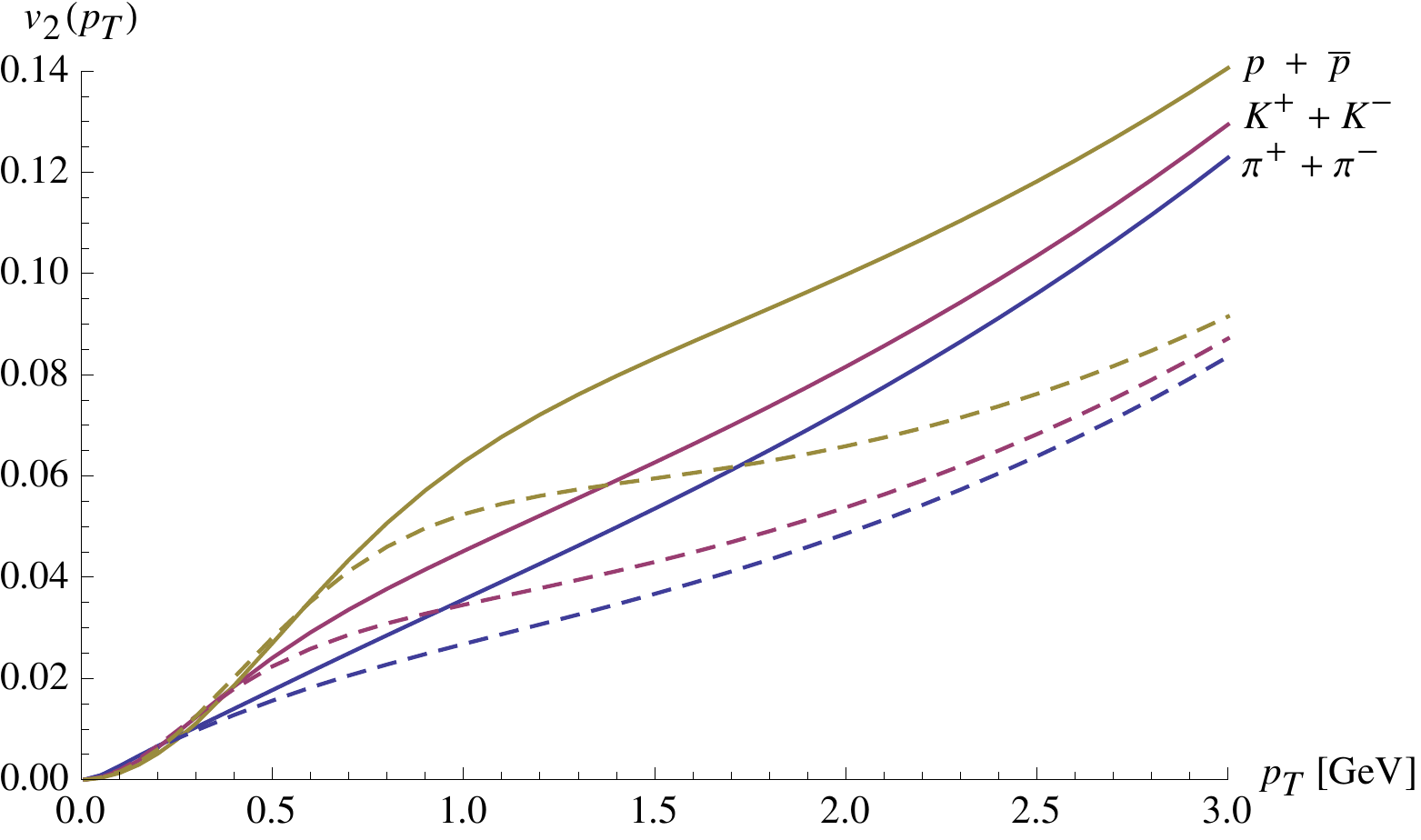}
\includegraphics[width=0.35\textwidth]{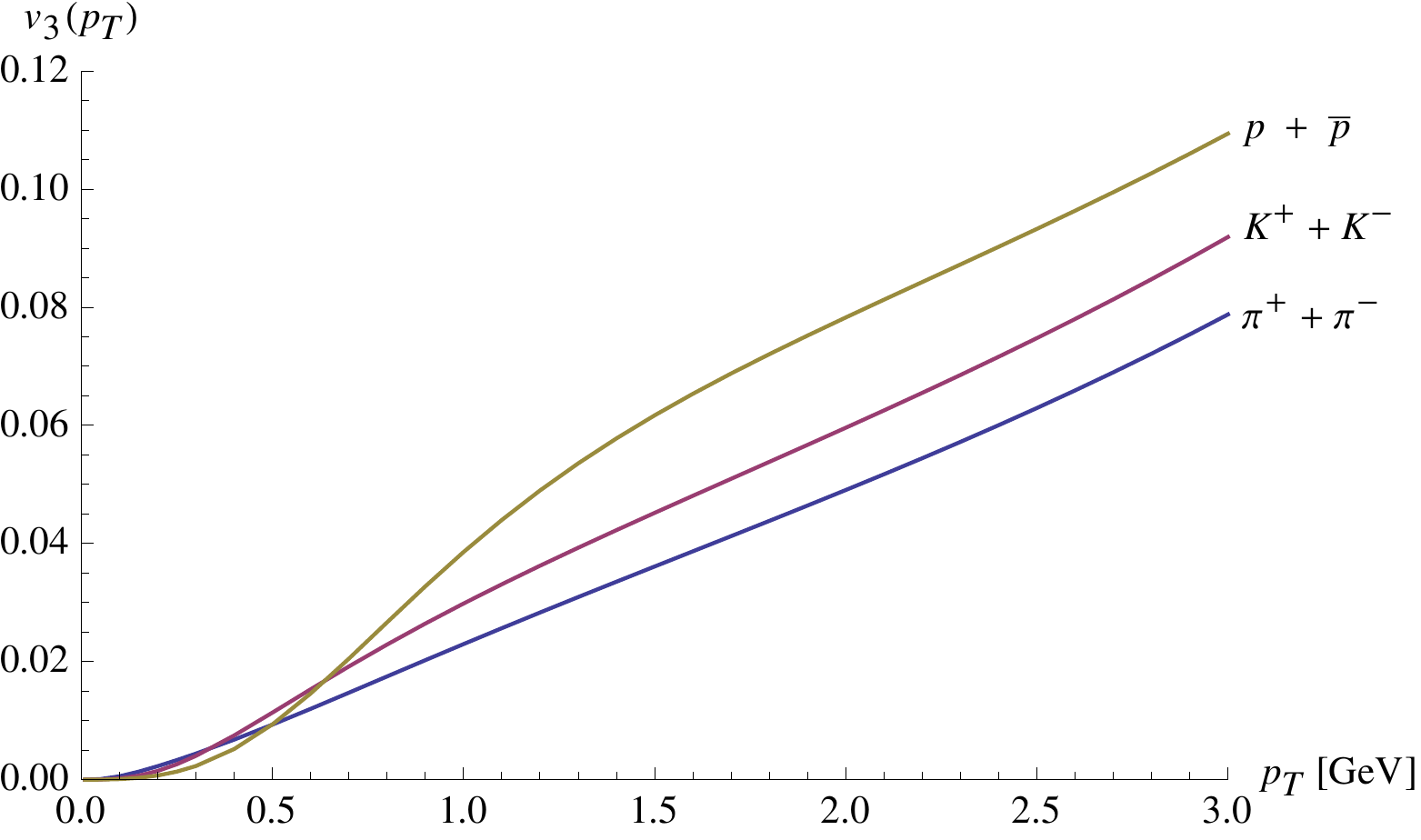}
\includegraphics[width=0.35\textwidth]{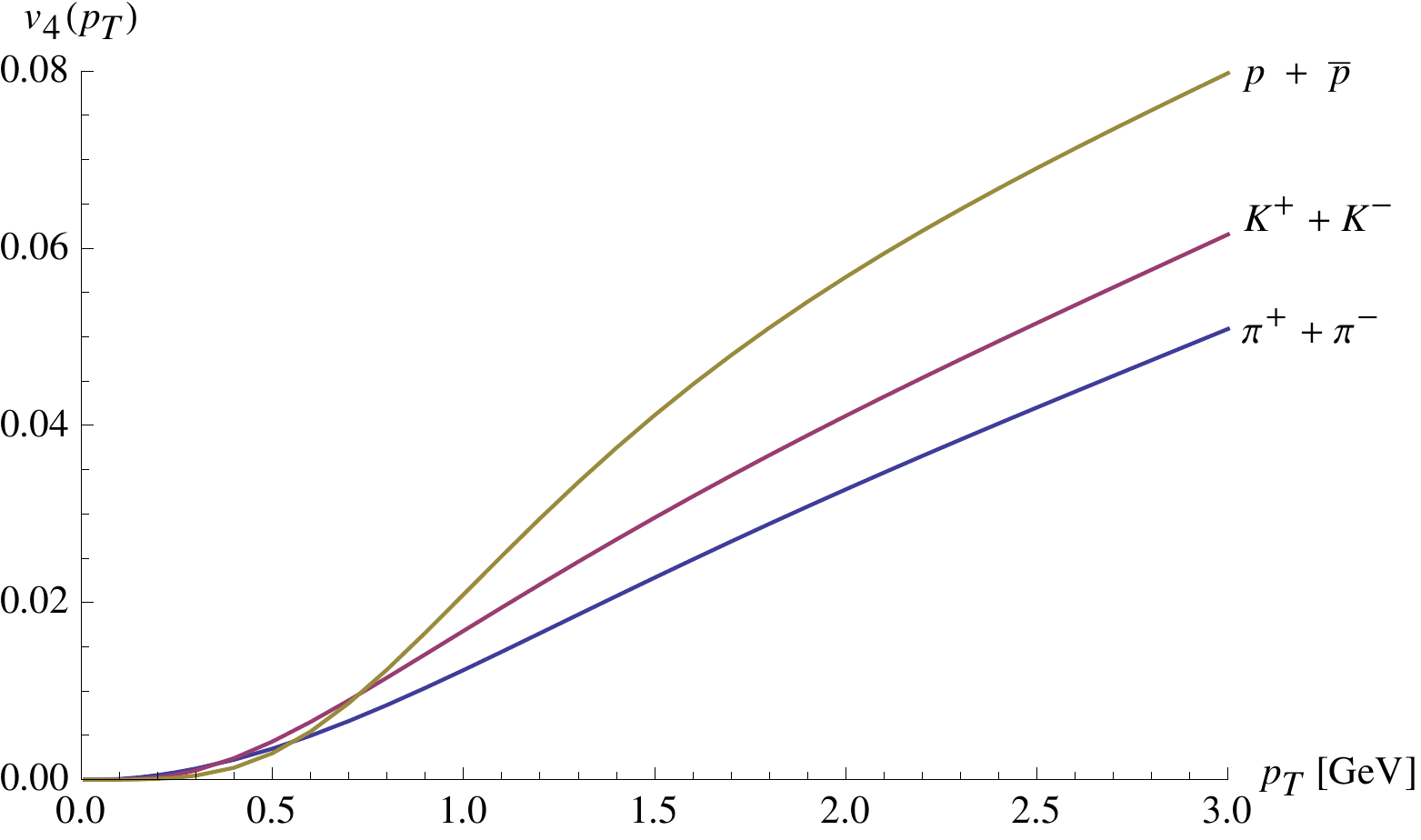}
\includegraphics[width=0.35\textwidth]{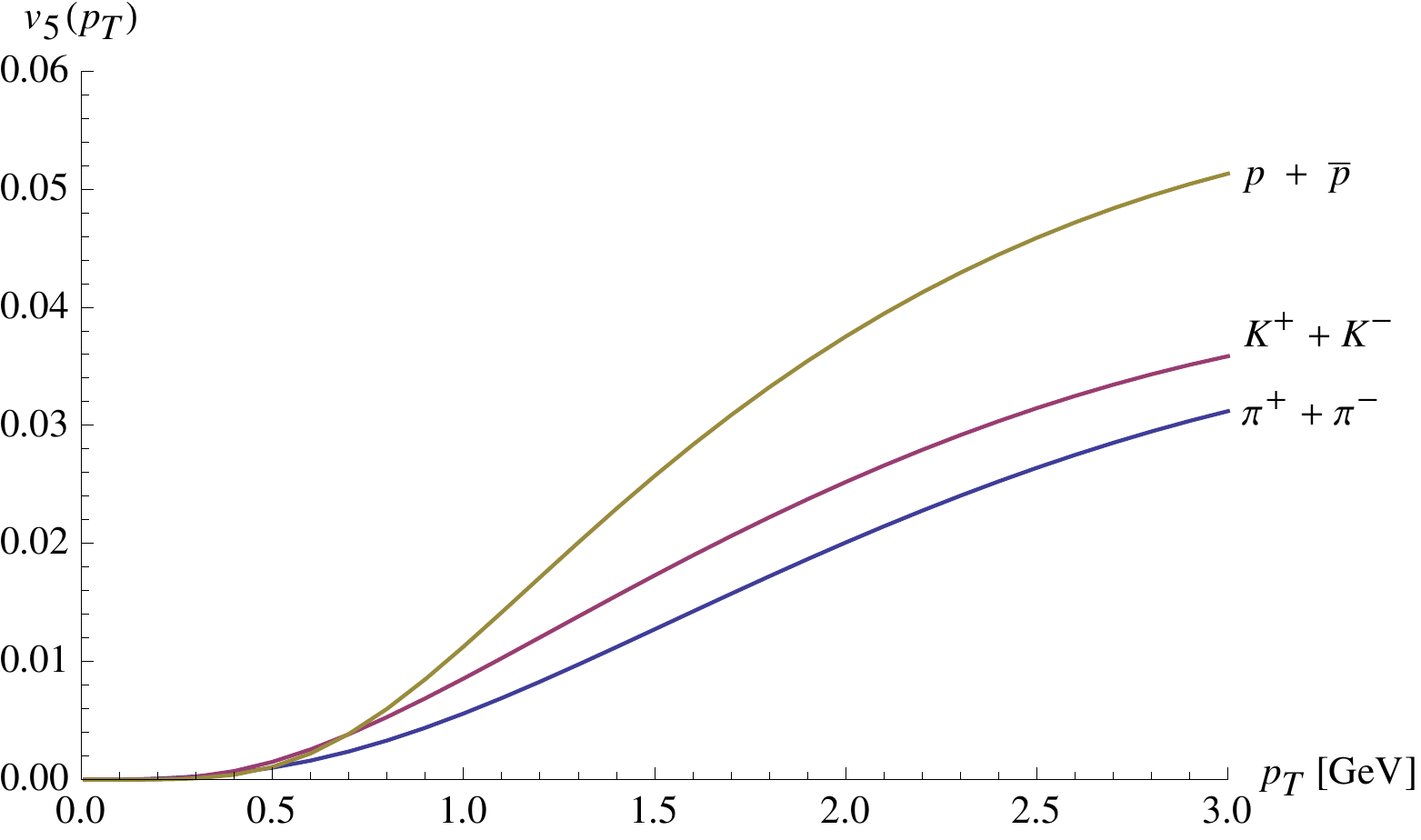}
\caption{The transverse momentum dependent $n$-th order flow coefficients
$v_n$, as in Fig.~\ref{fig4}, but calculated differentially for contributions from  
$\pi^\pm, K^\pm, p$ and $\bar p$  for central events. 
}
\label{fig5}
\end{figure}

The precise numerical values for $v_n(p_T)$ will depend in general on the weights and
correlations $\langle \tilde{w}_{l_1}^{(m)*}\, \tilde{w}_{l_2}^{(m)}\rangle$ of the different
fluctuating modes in the initial conditions, on the input parameters of the fluid dynamic
evolution, and on the treatment of rescattering effects and resonance decays after freeze-out.
The present study does not optimize input parameters and it does not account for physics
effects after freeze-out. Also, it is limited to the one set of fluctuating initial conditions
characterized in fig.\ \ref{fig2}. Within this non-optimized setting, we find a very reasonable
agreement with data for $v_2$, $v_3$, $v_4$ and $v_5$ in the range $p_T \leq 1$ GeV,
while experimental data  for $v_2$ in the range $1 \leq p_T \leq 3$ GeV lie significantly 
below our calculation. A full exploration of this significant, and other smaller 
discrepancies between data and calculation will require to optimize the input parameters 
which lies outside the scope of the present study. We note, however, that a 
simple mild rescaling of the weight of one single fluctuating mode in the initial conditions,
$ \tilde w^{(2)}_{1} \to 0.7 \, \tilde w^{(2)}_{1}$,  can improve agreement between
simulation results and data for $v_2$ over a much increased $p_T$-range. While
this curious observation clearly does not replace a full optimization of all input parameters,
it illustrates that mode-by-mode hydrodynamics offers the possibility of
``backward engineering'' of initial conditions: for any given dynamics and a set of data,
eqs.~\eqref{onespectrum} and \eqref{harmflow} allow to optimize the correlators 
$\langle \tilde w^{(m)}_{l_1} \tilde w^{(m)*}_{l_2} \rangle$, and thus the event 
distribution $p_{\tau_0}$. Further differential test of mode-by-mode fluid dyanamics will also
include the study of particle-identified flow harmonics. Fig.\ \ref{fig5} shows corresponding results
for pions, kaons and protons. Close inspection shows that the curves are ordered with the particle mass at small $p_T$ according to $v_m(p_T)^\text{Protons} < v_m(p_T)^\text{Kaons} < v_m(p_T)^\text{Pions}$, while for larger $p_T$ the ordering is reversed.

The proposed set-up may also be interesting in the context of 
the recently proposed  ``event shape engineering''~\cite{Schukraft:2012ah}. Namely, 
it allows easily for the calculation of event distributions  in one- and two-particle spectra 
from $p_{\tau_0}$, and it thus allows to study the relations between cuts on 
event distributions and cuts on initial conditions $p_{\tau_0}$.  Since it offers such possibilities,
we expect that the proposed fully differential treatment of fluctuations will become a
helpful tool used to fully exploit the experimental precision in heavy ion physics. 

\paragraph{Acknowledgements.}
S.\ F.\ acknowledges financial support by DFG under contract FL 736/1-1. We thank U. Heinz, M. Luzum, J.-Y. Ollitrault, H. Petersen and C. Shen for 
discussion.


\vskip -.5cm
\begin{thebibliography}{}
\bibitem{Alver:2010gr}
  B.~Alver and G.~Roland,
  Phys.\ Rev.\ C {\bf 81} (2010) 054905
   [Erratum-ibid.\ C {\bf 82} (2010) 039903].

\bibitem{Mishra:2007tw}
  A.~P.~Mishra, R.~K.~Mohapatra, P.~S.~Saumia and A.~M.~Srivastava,
  Phys.\ Rev.\ C {\bf 77} (2008) 064902.
  
\bibitem{Broniowski:2007ft}
  W.~Broniowski, P.~Bozek and M.~Rybczynski,
  Phys.\ Rev.\ C {\bf 76} (2007) 054905.
  
\bibitem{Sorensen:2008zk}
  P.~Sorensen [STAR Collaboration],
  J.\ Phys.\ G {\bf 35} (2008) 104102.
  
\bibitem{Takahashi:2009na}
  J.~Takahashi, B.~M.~Tavares, W.~L.~Qian, R.~Andrade, F.~Grassi, Y.~Hama, T.~Kodama and N.~Xu,
  Phys.\ Rev.\ Lett.\  {\bf 103} (2009) 242301.
  
\bibitem{Heinz:2013th}
  U.~W.~Heinz and R.~Snellings,
  arXiv:1301.2826 [nucl-th].
%

\bibitem{Gale:2013da}
  C.~Gale, S.~Jeon and B.~Schenke,
  Int.\ J.\ Mod.\ Phys.\ A {\bf 28} (2013) 1340011.

\bibitem{Qiu:2011iv}
  Z.~Qiu and U.~W.~Heinz,
  Phys.\ Rev.\ C {\bf 84} (2011) 024911.
  
\bibitem{Schenke:2011bn}
  B.~Schenke, S.~Jeon and C.~Gale,
  Phys.\ Rev.\ C {\bf 85} (2012) 024901.

\bibitem{Bhalerao:2011yg}
  R.~S.~Bhalerao, M.~Luzum and J.~-Y.~Ollitrault,
  Phys.\ Rev.\ C {\bf 84} (2011) 034910.
  
  
\bibitem{Schenke:2012wb}
  B.~Schenke, P.~Tribedy and R.~Venugopalan,
  Phys.\ Rev.\ Lett.\  {\bf 108} (2012) 252301.

\bibitem{Holopainen:2010gz}
  H.~Holopainen, H.~Niemi and K.~J.~Eskola,
  Phys.\ Rev.\ C {\bf 83} (2011) 034901.

\bibitem{Teaney:2010vd}
  D.~Teaney and L.~Yan,
  Phys.\ Rev.\ C {\bf 83} (2011) 064904.

\bibitem{Teaney:2012ke}
  D.~Teaney and L.~Yan,
  Phys.\ Rev.\ C {\bf 86} (2012) 044908.


\bibitem{Qiu:2011hf}
  Z.~Qiu, C.~Shen and U.~Heinz,
  Phys.\ Lett.\ B {\bf 707}, 151  (2012).

\bibitem{Gardim:2011xv}
  F.~G.~Gardim, F.~Grassi, M.~Luzum and J.~-Y.~Ollitrault,
  Phys.\ Rev.\ C {\bf 85} (2012) 024908.

\bibitem{Qian:2013nba}
  W.~-L.~Qian, P.~Mota, R.~Andrade, F.~Gardim, F.~Grassi, Y.~Hama and T.~Kodama,
  arXiv:1305.4673 [hep-ph].

\bibitem{Petersen:2012qc}
  H.~Petersen, R.~La Placa and S.~A.~Bass,
  J.\ Phys.\ G {\bf 39} (2012) 055102.

\bibitem{Deng:2011at}
  W.~-T.~Deng, Z.~Xu and C.~Greiner,
  Phys.\ Lett.\ B {\bf 711} (2012) 301.
  
\bibitem{Bhalerao:2011bp}
  R.~S.~Bhalerao, M.~Luzum and J.~-Y.~Ollitrault,
  Phys.\ Rev.\ C {\bf 84} (2011) 054901.

\bibitem{Gale:2012rq}
  C.~Gale, S.~Jeon, B.~Schenke, P.~Tribedy and R.~Venugopalan,
  Phys.\ Rev.\ Lett.\  {\bf 110} (2013) 012302.
   
\bibitem{Niemi:2012aj}
  H.~Niemi, G.~S.~Denicol, H.~Holopainen and P.~Huovinen,
  arXiv:1212.1008 [nucl-th].

\bibitem{Floerchinger:2013vua}
  S.~Floerchinger and U.~A.~Wiedemann,
  arXiv:1307.7611 [hep-ph].


  
\bibitem{Staig:2010pn}
  P.~Staig and E.~Shuryak,
  Phys.\ Rev.\ C {\bf 84} (2011) 034908.

\bibitem{Staig:2011wj}
  P.~Staig and E.~Shuryak,
  Phys.\ Rev.\ C {\bf 84} (2011) 044912.
  
\bibitem{Gubser:2010ui}
  S.~S.~Gubser and A.~Yarom,
  Nucl.\ Phys.\ B {\bf 846} (2011) 469.
  
\bibitem{Florchinger:2011qf}
  S.~Floerchinger and U.~A.~Wiedemann,
  JHEP {\bf 1111} (2011) 100.

\bibitem{ColemanSmith:2012ka}
  C.~E.~Coleman-Smith, H.~Petersen and R.~L.~Wolpert,
  arXiv:1204.5774 [hep-ph].
  
\bibitem{Springer:2012iz}
  T.~Springer and M.~Stephanov,
  Nucl.\ Phys.\ A904-905 {\bf 2013} (2013) 1027c
  [arXiv:1210.5179 [nucl-th]].
 
\bibitem{Lemoine}
  D.~Lemoine, J.\ Chem.\ Phys.\ {\bf 101}, 3936 (1994).



\bibitem{Abelev:2012wca}
  B.~Abelev {\it et al.}  [ALICE Collaboration],
  Phys.\ Rev.\ Lett.\  {\bf 109}, 252301 (2012).

\bibitem{ALICE:2011ab}
  K.~Aamodt {\it et al.}  [ALICE Collaboration],
  Phys.\ Rev.\ Lett.\  {\bf 107}, 032301 (2011).
  
\bibitem{Schukraft:2012ah}
  J.~Schukraft, A.~Timmins and S.~A.~Voloshin,
  Phys.\ Lett.\ B {\bf 719} (2013) 394.
  
  
\bibitem{Teaney:2003kp}
  D.~Teaney,
  Phys.\ Rev.\ C {\bf 68} (2003) 034913.
  
  
\end{thebibliography}
\end{document}